\documentclass{article}
\usepackage{graphicx} \usepackage{amssymb}

\suppressfloats[t]

\begin{document} %\large

\noindent
{\large\bf Macromechanical behavior of oxide nanopowders during compaction processes}\\[2ex]

\noindent
Boltachev G.Sh.$^{1(*)}$, Volkov N.B.$^1$, Kochurin E.A.$^1$,
  Maximenko A.L.$^2$, Shtern M.B.$^2$, Kirkova E.G.$^2$\\
$^1$Institute of Electrophysics, Ural Branch of Russian Academy of Sciences,
    Amundsen Street 106, 620016, Ekaterinburg, Russia\\
$^2$Frantsevich Institute for Problems of Materials Science, NAS Ukraine,
    Krzhizhanovsky Street 3, 03680, Kyev, Ukraine\\
$^{*}$E-mail: grey@iep.uran.ru, tel.: +7(343)2678776

\vspace{5mm}

\noindent {\bf Abstract.}
Two granular systems (I and II) corresponding oxide nanopowders having different
  agglomeration tendency are simulated by the granular dynamics method.
The particle size is 10 nanometer.
The interaction of particles involves the elastic forces of repulsion, the tangential
  forces of "friction", the dispersion forces of attraction, and in the case of II system
  the opportunity of creation/destruction of hard bonds of chemical nature.
The processes of the uniaxial compaction, the biaxial (radial) one, the isotropic one,
  the compaction combined with shear deformation as well as the simple shear deformation
  are studied.
The effect of the positive dilatancy is found out in the processes of shear deformation.
The loading surfaces of nanopowders are constructed in the space of stress tensor
  invariants, i.e., the hydrostatic pressure and the deviator intensity.
It is revealed that the form of the loading surfaces is similar to an ellipse, which is
  shifted along the hydrostatic axis to compressive pressures.
The associated flow rule is analyzed.
The nonorthogonality of the deformation vectors to the loading surface is established
  in the both systems modeled.

\vspace{3mm}


\vspace{5mm}

{\bf I. Introduction}

At present time high hopes concerning the development of promising structural and functional
  materials are pinned on the production and the investigation of nanostructured ceramics
  based on different oxides, for example, Al$_2$O$_3$, ZrO$_2$, Y$_2$O$_3$, YSZ, and so on
  \cite{l.Siegel,l.Iv05,l.Ivan06,l.Iv07,l.Kayg07}.
The powder metallurgy methods, which include such stages as the nanopowder production,
  compaction, and sintering, are the most debugged and fruitful methods for manufacture
  of the nanosized ceramic materials
  \cite{l.Iv05,l.Ivan06,l.Iv07,l.Kayg07,l.Shtern,l.Nova}.
The nanostructure preservation during the sintering stage needs the use of reduced temperatures
  that makes great demands of compacts prepared in previous stage
  \cite{l.Iv05,l.Ivan06,l.Iv07,l.Kayg07}.
Generally, the compacts must be uniform and have large density.
On the other hand, the size effect in the compaction processes is known.
It is harder to compact the nanopowders as compared to the powders consisting of larger
  size particles in view of the presence of relatively large adhesion forces
  \cite{l.Filon,l.Vass,l.Zhao,l.Saha}, which resilt from dispersion forces of attraction
  \cite{l.Balak,l.ZTF11,l.Runano,l.PRE13}.
To overcome the strong adhesion of nanopowders it is necessary to use very high pressures,
  which can turned out beyond the ultimate strength of experimental setup
  \cite{l.Bol13c}.
Such high pressures are achieved, for example, in the processes of magnetic pulsed
  compaction owing to the application of inertial effects \cite{l.Nova}.
In spite of the considerable experimental progress in this line 
  \cite{l.Iv05,l.Ivan06,l.Iv07,l.Kayg07}, it can be claimed that the further progress
  is impeded by the lack of theoretical description of nanosized powders.

As a rule, present-day theoretical approaches to nanopowder description are the
  application of principles, which have developed for larger particle powders and
  do not take into account the peculiarities of new subjects of inquiry.
So, for example, the theory of the plastically hardening porous body
  \cite{l.Shtern,l.Nova,l.Max08} and the related models of sintering and hot-pressing
  \cite{l.Skor,l.Olev04} appear to be the powerful tools for the theoretical analysis.
On the whole, the results of such models have been verified experimentally for the processes
  of the cold compaction of the powders consisting of micron- (or larger) size particles
  \cite{l.Shtern,l.Skor,l.Olev04}.
In the case of the cold compaction the plasticity theory of the materials having the volume
  compressibility has been applied.
Here, as a rule, they suppose that the behavior of bodies being compacted can be describe
  in terms of the yield stress and the deformation hardening both on a macroscopic scale
  and on a particle scale.
In the case of nanosized powders, especially of oxide ones, where the particles cannot
  deform plastically \cite{l.Filon,l.Gryaznov}, such conceptions as the yield stress
  or hardening takes very relative meaning.
As a result, the adaptability of the continuous theory requires additional
  vindications at the least.
On the other hand, the sophisticated analysis and revision of continuous theories
  are impossible without a lot of laborious experiments, which should provide 
  additional information on subjects of inquiry (nanosized powders).

The granular dynamics method (or discrete element method) suggested for the first time
  by Cundall and Strack \cite{l.Cund} and being developed strongly at present time 
\cite{l.Balak,l.ZTF11,l.Runano,l.PRE13,l.Agno,l.GRC07,l.GRC08,l.Zhu08,l.Ludi08,l.Salo09,l.Teuf09,l.Piz,l.Yang13}
  is a convenient alternative to real and laborious experiments and is a very promising
  approach, which allows finding the detailed information on powder systems.
Due to a great extent of the sphericity and the non-deformability (strength) of individual
  particles of oxide nanopowders being studied, the granular dynamics method is particularly
  attractive and promising tool of the theoretical analysis.
The low plastic deformability of particles with size of 10 -- 100 nm is connected with their
  defects free: the dislocations are ejected from the particles by the high "image"{} forces
  \cite{l.Gryaznov}.
This kind of particles (especially oxide nanoparticles) is elastically deformed with
  unloading shape recovery.
Within the bounds of the granular dynamics this fact allows us to avoid the difficulties
  related to accounting the plastic deformation of individual particles.
Thus we emphasize that the simulation presented in this study can be used to describe
  the behavior of granular systems where the appearance of the plastic shape change of
  particles can be neglected (like oxide nanopowders).

This paper is a continuation of the investigations 
  \cite{l.ZTF11,l.Runano,l.PRE13}.
References~\cite{l.ZTF11,l.Runano} were devoted to 2D simulations.
In Ref.~\cite{l.PRE13} the discrete three-dimensional model of oxide nanopowders was developed.
Using this model we succeeded in reproducing the known properties of different
  alumina-based nanopowders.
In this study two monodisperse systems corresponding oxide nanopowders having weak
  (system I) and strong (system II) agglomeration tendency are simulated.
The numerical experiments are formulated in 3D geometry: the particles of the model
  systems have a spherical shape with diameter $d=10$ nm and possess both translation
  and rotation degrees of freedom.
In addition to the contact interaction laws commonly used, the interaction of particles
  involves the dispersion forces of attraction and in the case of II system the
  opportunity of creation/destruction of hard interparticle bonds.
These bonds appear as a result of hard pressing force between particles which is initiated
  or by action of high dispersion interaction or by compaction process, Ref.~\cite{l.PRE13}.

\vspace{5mm}

{\bf II. Calculation procedure}

The model cell is a right-angle prism with sizes of $x_{cell}$, $y_{cell}$ and $z_{cell}$.
For initial packing generation the algorithm defined in Ref.~\cite{l.PRE13} is used, which
  allows us to create isotropic and uniform structures in a form of the connected 3D-periodic
  cluster with chains thickness of 2 particles.
The total number of particles $N_p$ is 8000, the initial density $\rho_0$ is 0.24.
The density $\rho$ is a relative volume of the solid phase, that is
  $\rho=(\pi/6)N_p d^3/V_{cell}$, where $V_{cell}$ is the model cell volume.
The periodic boundary conditions are used on all the faces of the cell.
The system deformation is carried out by simultaneous changes of selected sizes of the
  model cell and proportional rescaling of the appropriate coordinates of all the particles.
After every act of deforming the new equilibrium location of the particles is determined.
This procedure corresponds to the quasi-static conditions of the powder compaction.

The stress tensor $\sigma_{ij}$ averaged over the model cell was calculated by the known
  expression \cite{l.Agno,l.GRC07,l.GRC08}
\begin{equation}
\sigma_{ij} = \frac{-1}{V_{cell}} \sum_{k<l} f_i^{(kl)} r_j^{(kl)},
\label{sigij}
\end{equation}
  where the summation is performed over all the pairs of the interacting particles $k$ and
  $l$; $\vec f^{(kl)}$ is the total force effecting the particle $k$ from the particle $l$;
  $\vec r^{(kl)}$ is the vector connecting the centers of the considered particles.
The interaction forces between particles are described by the relationships
  \cite{l.PRE13}:
\begin{equation}
f_a(r) = \frac{\pi^2}{3}
  \frac{\left(nd_0^3\right)^2 \varepsilon d^6}{(r+\alpha d_0)^3\left[(r+\alpha d_0)^2-d^2\right]^2}~,
\ \ \ \ \  r = |\vec r|~,
\label{fah}
\end{equation}
\begin{equation}
\frac{f_e(r)}{Ed^2} = \frac{(h/d)^{3/2}}{3(1-\nu^2)} - \frac{\pi}{4}
  \frac{k_r(1-\nu)}{(1-2\nu)(1+\nu)}
  \left[ \frac{h}{d} + \ln\left(1-\frac{h}{d}\right) \right]~, \ \ \ \ 
h = d-r~,
\label{feh}
\end{equation}
\begin{equation}
f_t(\delta) = \min\left\{ \frac{4Ea\delta}{(2-\nu)(1+\nu)}\ ; \ \mu f_e \
  ; \ \pi a^2 \sigma_b  \right\}~, \ \ \ \
a = \frac{\sqrt{hd}}{2}~,
\label{fMi}
\end{equation}
\begin{equation}
M_p(\theta_p) = \min\left\{ \frac{8Ea^3}{3(1+\nu)}\ \theta_p ; \ \mu M(a)\ ;
  \ \frac{\pi}{2} a^3 \sigma_b  \right\}~, \ \ \ \ \
M(a) = -2\pi\int_0^a \sigma_n(r) r^2 dr~,
\label{Mpiv}
\end{equation}
\begin{equation}
M_{r}(\theta_r) = \min \left\{  \frac43 \frac{Ea^3}{1-\nu^2} \ \theta_r~;
  \ \frac13\ a f_{e} \right\}~.
\label{Mroll}
\end{equation}
Here: the modified Hamaker formula (\ref{fah}) determines the dispersive attraction force $f_a$;
  the modified Hertz law (\ref{feh}) gives the elastic repulsive force $f_e$;
  the linearized Cattaneo--Mindlin law (\ref{fMi}) defines tangential interaction of
  the pressed particles (forces of "friction"{}); the linearized J\"ager law (\ref{Mpiv})
  (or the Reissner-Sagoci law, Ref.~\cite{l.Reiss}) describes the surface forces moment $M_p$
  arising under the rotation of the pressed particles about the contact axis at the angle
  $\theta_p$; the Lurie law (\ref{Mroll}) is the surface forces moment $M_r$ arising under
  the bending of the contact axis at the angle $\theta_r$ (only in the presence of hard
  interparticle bond; Ref.~\cite{l.Lurie}, page 272, Eq.~(4.5)).
In the Eqs.~(\ref{fah})--(\ref{Mroll}): $\varepsilon$ and $d_0$ are the power and dimensional parameters
  of the intermolecular forces; $\alpha$ is the coefficient determining the minimum
  gap between the contacting particles (at $r=d$) and thus specifying the maximum
  force of attraction ($f_{a,\max}=f_a(d)$); $E$ and $\nu$ are the Young modulus and
  the Poisson ratio of the particles; $\delta$ is the tangential displacement of the contact
  area; $a$ is the contact area radius; $\mu$ is the friction coefficient; $\sigma_b$ is the
  critical shear stress defining the material shear strength; $\sigma_n$ is the normal
  stress on the contact surface.

Creation/destruction of the hard interparticle bond is described by the parameter
  $\Delta r_{ch}$ characterizing the necessary interparticle pressing force \cite{l.PRE13}.
It is assumed that the decrease of distance $r$ between the particle centers up to the value
  $r_{\min}\le d-\Delta r_{ch}$ initiates the hard bond creation.
After the creation of the hard interparticle bond the further compression (as $r$ decreases)
  continues to obey the elastic interaction law (\ref{feh}) and at the extension (as $r$
  increases) we have linear relationship between the force $f_e$ and the distance $r$ up to
  the distance $r'=r_{\min}+\Delta r_{ch}$.
For distances $r>r'$ the partial contact destruction is introduced and described by the increase of
  the parameter $r_{\min}$ so that the difference $r-r_{\min}$ should remain equal to its
  maximum value $\Delta r_{ch}$.
The total destruction of the interparticle contact occurred at the extension up to the
  distance $r=d$.
The restrictions in Eqs.~(\ref{fMi}) and (\ref{Mpiv}) related to the friction
  coefficient $\mu$ are removed with creation of the hard interparticle bond.

Alumina in the $\alpha$-phase is implied as the particle material for which we accept:
  $E=382$ GPa, $\nu=0.25$, $nd_0^3=\sqrt2$, $d_0=0.392$ nm; $\varepsilon=1224k_B$,
  $\sigma_b=0.018E$ \cite{l.PRE13}.
Simulations are performed for two model monodisperse systems with particle diameter
  $d=10$ nm.
The model system I with the parameters $\alpha=0.37$ and $\mu=0.13$ is characterized by
  the absence of the hard interparticle bonds (an unreal high value equal $d$ is assigned
  to the parameter $\Delta r_{ch}$) and corresponds to the weakly agglomerating powder
  of Ref.~\cite{l.PRE13}.
In the model system II with the parameters $\alpha=0.24$, $\mu=0.10$ and
  $\Delta r_{ch}=0.008d$ the opportunity of the hard interparticle bond creation is taken
  into account.
This system corresponds to the strongly agglomerating powder of Ref.~\cite{l.Runano}.
Larger value of the interparticle gap ($\alpha$) and the suppression of agglomeration
  in I system are due to the adsorbate presence on the particle surface. 

We carried out the computer experiments for the following processes: 

A. The uniform (triaxial) compression: on the every deformation step of the model cell all
  its sizes decreased simultaneously by 0.1\% from the current values.
In this case, the strain rate tensor $e_{ij}$ and stress tensor $\sigma_{ij}$ are
  spherical tensors, that is $e_{ij}=(e/3)\delta_{ij}$, $\sigma_{ij}=\sigma_z\delta_{ij}$
  ($\delta_{ij}$ is the unit tensor, $e=\mbox{Tr}(e_{ij})$).

B. The biaxial compression along the axes $Oy$ and $Oz$.
   The strain rate and stress tensors are characterized by the values $e_{xx}=0$,
   $e_{yy}=e_{zz}=e/2$; $\sigma_{xx} \not = \sigma_{yy}=\sigma_{zz}$.
   For deviator intensity of the strain rate tensor $\gamma=\sqrt{\gamma_{ij}\gamma_{ji}}$
   ($\gamma_{ij} = e_{ij} - (e/3)\delta_{ij}$) we have $\gamma=|e|\sqrt{1/6}$, and for
   deviator intensity of the stress tensor $\tau=\sqrt{\tau_{ij}\tau_{ji}}$
   ($\tau_{ij}=\sigma_{ij}-\delta_{ij}\mbox{ Tr}(\sigma_{ij}) / 3$) we have
   $\tau=|\sigma_{xx}-\sigma_{zz}|\sqrt{2/3}$.

C. The uniaxial compression along the axis $Oz$: $e_{xx}=e_{yy}=0$, $e_{zz}=e$; $\sigma_{zz}<0$,
   $\sigma_{xx}=\sigma_{yy}=\sigma_t$; $\gamma=|e|\sqrt{2/3}$,
   $\tau=|\sigma_t-\sigma_{zz}|\sqrt{2/3}$.

D. The $z$-compression combined with application of the shear deformation:
   the simultaneous compression in the $Oz$ direction (the value $z_{cell}$ decreased by 0.1\% from the current value)
   and the extension in the $Oy$ direction (the value $y_{cell}$ increased by 0.05\% from
   the current value) were carried out on the every deformation step.
   The strain rate tensor is characterized by the values: $e_{xx}=0$, $e_{yy}=-e$,
   $e_{zz}=2e$; $\gamma=|e|\sqrt{14/3}$.

E. The shear deformation of the model cell at constant volume.
   Here the value $z_{cell}$ decreased and the value $y_{cell}$ increased simultaneously
   (by 0.1\% from the current values) on the every deformation step:
   $e_{xx}=0$, $e_{yy}=-e_{zz}$; $e=0$, $\gamma=|e_{yy}|\sqrt2$.

\vspace{5mm}

{\bf III. Extraction of "elasto-reversible"{} part}

\begin{figure*}  \hfill
\includegraphics*[width=0.49\textwidth, height=!]{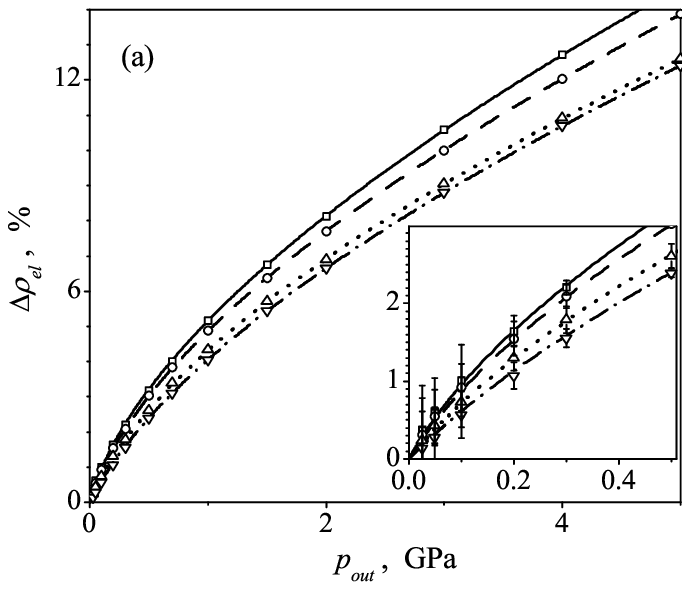} \hfill
\includegraphics*[width=0.49\textwidth, height=!]{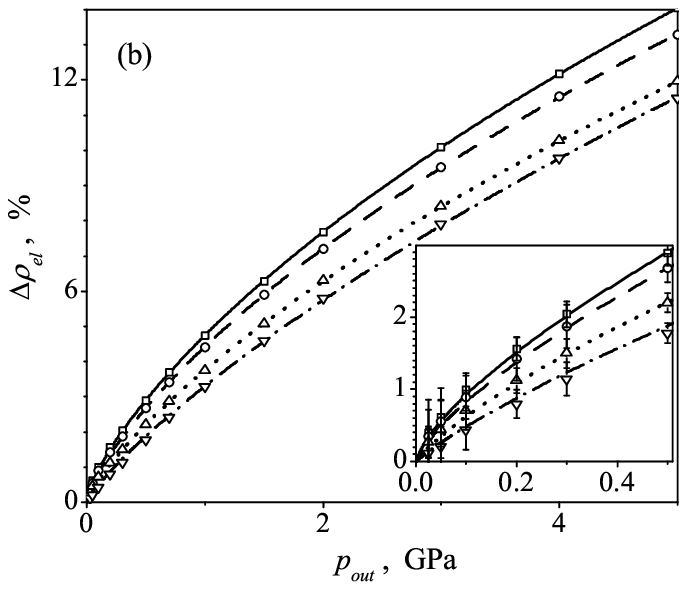} \hfill
\caption{ \label{f.Gdroep} 
  Density change during elastic unloading ($\Delta\rho_{el}=\rho_p-\rho_u$) versus
  external pressure $p_{out}=-\sigma_{zz}$ for systems I (a) and II (b).
  Points correspond to the simulation results and lines correspond to the approximations
  by Eq.~(\ref{proel}) for the processes A (solid lines, squares), B (dashed lines,
  circles), C (dotted lines, upper triangles), and D (dash-and-dot lines, lower triangles).
  The insets show the low pressure zones in an enlarged scale.}
\end{figure*}

In the calculations with compression (the processes A--D) the model cell compaction was
  carried out up to the prescribed level $p_{max}$ of the external load along the axis $Oz$.
Then unloading of the model cell was performed, during which the cell was expanded in all
  directions with the rates proportional to the corresponding stresses:
  $e_{ii}\propto\sigma_{ii}$.
This stage (the elastic unloading \cite{l.Salo09}) is characterized by the density change
  $\Delta\rho_{el}$.
Apparently, besides purely elastic unloading of interparticle contacts the pressure unloading
  is also accompanied by irreversible processes of relative particle displacement.
So naming these stages "elastic"{} is rather relative and implies only that the elastic
  processes here dominate.
The elastic unloading calculations were performed for the values $p_{max}=0.025$, 0.05, 0.1,
  0.2, 0.3, 0.5, 0.7, 1, 1.5, 2, 3, 4 and 5 GPa.
The relationship between the applied external pressure $p_{out} = -\sigma_{zz}$ and the
  value $\Delta\rho_{el}$ was approximated by the expressions
\begin{equation}
p_{out} = s_1 \Delta\rho_{el} + s_2 \Delta\rho_{el}^2 +
          \frac{s_3 \Delta\rho_{el}}{(s_4+\Delta\rho_{el})^2}.
\label{proel}
\end{equation}
Approximation coefficients for system I: $s_1=24.3$, $s_2=104.1$ (the process A);
  $s_1=25.3$, $s_2=118.4$ (the process B); $s_1=28.1$, $s_2=140.7$ (the process C);
  $s_1=30.8$, $s_2=126.2$ (the process D); and $s_3=-0.65$, $s_4=0.2$ for all the processes.
Approximation coefficients for system II: $s_1=16.0$, $s_2=142.0$ (the process A);
  $s_1=17.7$, $s_2=151.9$ (the process B); $s_1=22.7$, $s_2=163.6$ (the process C);
  $s_1=28.3$, $s_2=135.3$ (the process D); and $s_3=-0.01$, $s_4=0.03$ for all the processes.
Figure \ref{f.Gdroep} shows that the suggested approximations describe the calculated data
  with the error, which is not more than the statistical averaging error
  (each of the calculations consisted of 10 independent computer experiments).
The inverse relations $\Delta\rho_{el}(p_{out})$ presented in the figure demonstrate
  nonlinear behavior, which appears with the greatest distinction for system II in the
  low pressure zones.
On the whole, presence of the hard bonds in system II leads to rather smaller values
  $\Delta\rho_{el}$ that relates to suppression of the irreversible processes of
  particle sliding (rearrangement) during the elastic unloading stage.
It should be noticed, however, that the differences are not large.
Thus, for the uniform compression processes (A) at the pressure $p_{out}=1$ GPa we have
  $\Delta\rho_{el}=5.2$\% in the system I and 4.7\% in the system II.

\begin{figure*}  \hfill
\includegraphics*[width=0.49\textwidth, height=!]{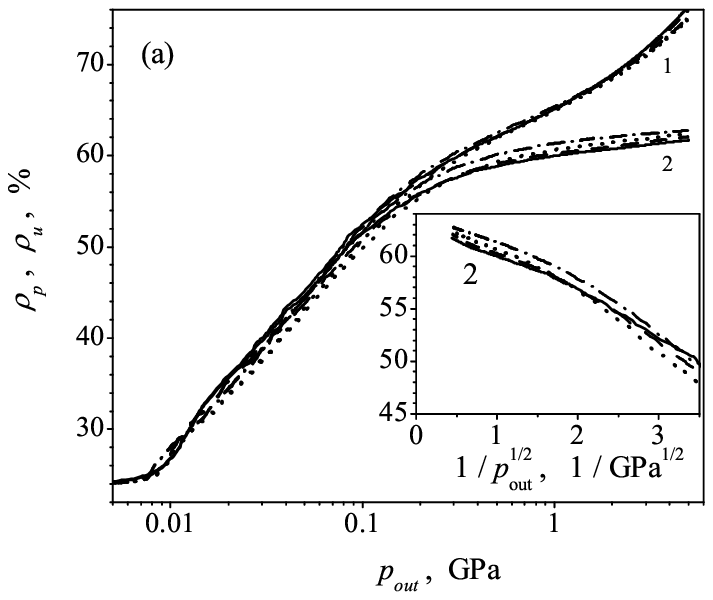} \hfill
\includegraphics*[width=0.49\textwidth, height=!]{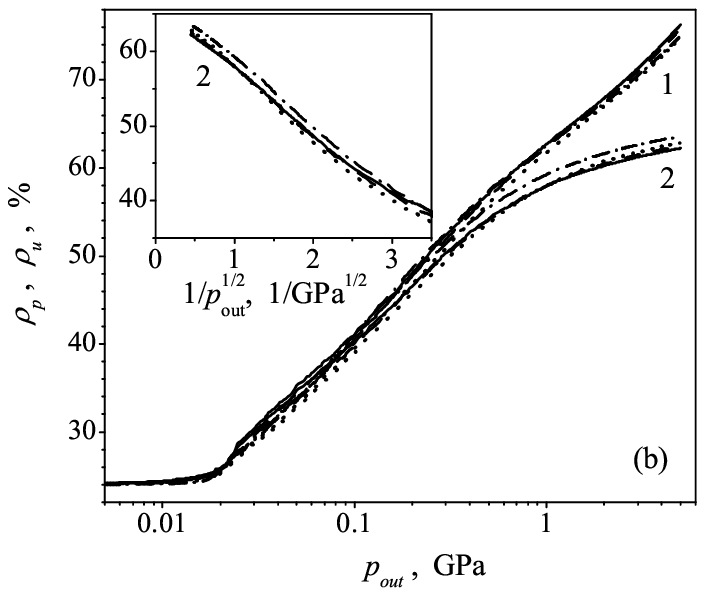} \hfill
\caption{ \label{f.Group}
  Density $\rho_p$ (under pressure) /1/ and unloading density $\rho_u$ /2/ versus
  external pressure $p_{out}$ for the processes A--D (lines are marked by the same symbols
  as in Fig.~1) in systems I (a) and II (b).
  The insets show the unloading density versus value of $p_{out}^{-1/2}$.}
\end{figure*}

Obtained relations (\ref{proel}) allow us to extract elasto-reversible part ($\Delta\rho_{el}$)
  from the total strain of the model cell and to isolate irreversible (plastic) component
  which is characterized by the material unloading density $\rho_u$.
The unloading density dependences on the external pressure for the analyzed systems are
  represented in Fig. \ref{f.Group}.
For comparison the initial dependences $\rho_p(p_{out})$ with the elastic contribution
  ($\rho_p = \rho_u + \Delta\rho_{el}$) are also represented there.
Interestingly enough that in the both systems the dependences $\rho(p)$ corresponding to
  the different processes (A--D) are rather close to each other and the maximum difference
  of the unloading densities is within 1-2\%.
On the whole, absence of the hard interparticle bonds in system I leads to its higher
  compactability: the curves $\rho_p(p_{out})$ and $\rho_u(p_{out})$ pass above for the system I.
But in the high pressure zone the curve $\rho_p(p_{out})$ differences become insignificant
  and large values of the elastic unloading ($\Delta\rho_{el}$) for the system I lead to a slight
  decrease of the unloading density as compared with system II.
For example, in the case of the uniform compression at the pressure $p_{out}=5$ GPa
  we have $\rho_p=76.4$\%, $\rho_u=61.7$\% in system I and $\rho_p=76.3$\%, $\rho_u=62.3$\%
  in system II.
In the hypothetical limit of the unrestrictedly large pressures ($p\to\infty$) the powder
  unloading density is about 65\% (see insertions in Fig. \ref{f.Group}).
The average coordination number $k$ of the simulated systems after unloading from
  $p_{\max}=5$ GPa is in the range 6.2--6.4 for the processes A--D.
Thus, it can be concluded that after the unloading from high values of the external pressure
  ($p_{out}>5$ GPa) the systems being studied are close to the random close packing (RCP)
  state, for which $\rho\simeq 0.64$ and $k\simeq 6$.

Qualitatively the dependences $\rho_u(p_{out})$ for systems I and II demonstrate similarities.
On the curves $\rho_u(p_{out})$ we can identify three qualitative stages which are
  known from experiments \cite{l.Castel,l.Valv07} and two-dimensional computer simulations
  \cite{l.GRC08} of micron-size powder compaction.
In the first stage the density weakly depends on the pressure.
Here the external loading is insufficient to overcome initial particle bonds.
This stage corresponds to the pressure $p_{out}<9$ MPa in system I and $p_{out}<20$ MPa
  in system II.
An active densification by the law $\Delta\rho \propto \ln(p_{out})$ is observed in the second
  stage.
Here the basic processes of particle rearrangement take place.
In the third stage when $p_{out}\gtrsim 200$ MPa in system I and $p_{out}\gtrsim 500$ MPa
  in system II the powder unloading density gradually reaches some maximum value
  $\rho_{u,\max}$ about 65\%.
The calculations show that here we have $p_{out} \propto 1/(\rho_{u,\max} - \rho_u)^2$
  (see insertions in Fig. \ref{f.Group}).
It should be noted that in two-dimensional structure simulations the respective power exponent
  is equal to one \cite{l.GRC08,l.PM12}, which allows us to suppose in the general case
  $p_{out} \propto 1/(\rho_{u,\max} - \rho_u)^{D-1}$, where $D$ is the space dimension.

\vspace{5mm}

{\bf IV. Shear deformation at constant volume}

\begin{table}[tb]
\centering
\caption{Calculation parameters for simulations of the model cell shear deformation
  as well as the densities for systems I and II at the shear stages.}
\vspace{3mm}
\begin{tabular}{|c|c|c|c|c|c|c|}
\hline
No. & $n_z$ & $L_n/d$ & $L_0/d$ & $N_p$ & $I:\ \rho$,\% & $II:\ \rho$,\% \\  \hline
  1 & 4 & 14 & 12 & 6000 & 36.1 & 35.6 \\
  2 & 3 & 14 & 12 & 5000 & 37.6 & 37.2 \\
  3 & 3 & 14 & 12 & 6000 & 45.0 & 44.3 \\
  4 & 3 & 14 & 12 & 7000 & 52.1 & 51.0 \\
  5 & 3 & 14 & 12 & 8000 & 58.1 & 56.9 \\
  6 & 3 & 14 & 12 & 9000 & 60.5 & 60.4 \\
  7 & 2 & 15 & 13 & 5000 & 39.5 & 38.9 \\
  8 & 2 & 15 & 13 & 6000 & 47.1 & 46.4 \\
  9 & 2 & 15 & 13 & 7000 & 54.3 & 53.5 \\
 10 & 2 & 15 & 13 & 8000 & 59.5 & 58.4 \\
 11 & 2 & 18 & 15 & 8000 & 41.1 & 40.3 \\
\hline
\end{tabular}
\end{table}

In simulations of the shear deformation the initial sizes of the model cell were determined
  equal to $x_{cell}=y_{cell}=L_n$, $z_{cell}= (n_z+1)L_n$ ($n_z\ge 2$).
At first, the uniform compression of the model cell was carried out up to the specified
  sizes ($L_n\to L_0$).
This step allowed us to obtain the powder structure with the necessary initial density $\rho_0$.
Then the unloading (not large reverse cell expansion) was performed.
After that, the shear deformation process was carried out: the model cell was compressed
  along the axis $Oz$ and elongated along the axis $Oy$ so that the volume remained constant.
The process was stopped when the values $z_{cell}=L_0$ and $y_{cell}=(n_z+1)L_0$
  were reached.
At last, the re-unloading till zero stresses was carried out.

These calculations have been performed for 11 parameter sets ($n_z$, $L_n$, $L_0$, $N_p$)
  which are listed in Table 1.
In the first unloading stage (after the initial uniform compression of the cell) the
  significant density decrease is observed, and it is different for analyzed systems I and II.
Obtained densities of these systems, characterizing the shear deformation stages
  at constant volume (at constant density), are also represented in Table 1.
Density changes in the final unloading stages are negligibly small.

\begin{figure}[tb]  \hfill
\includegraphics*[width=0.49\textwidth, height=!]{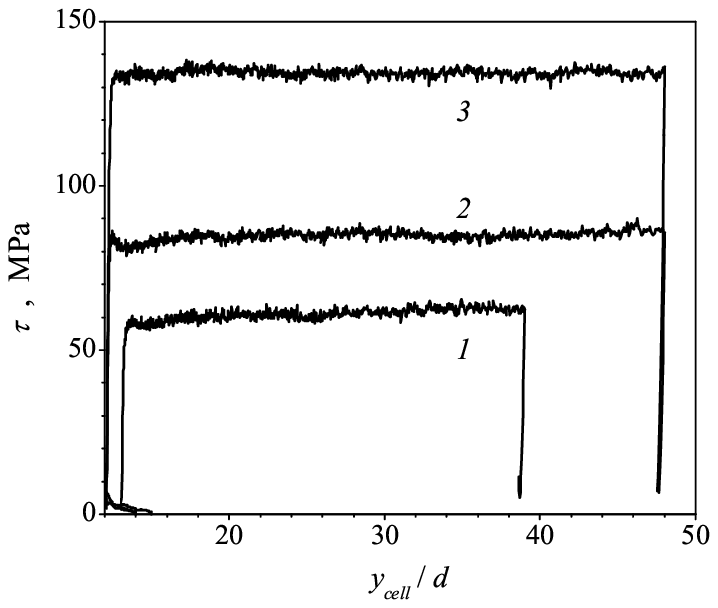}  \hfill
\includegraphics*[width=0.49\textwidth, height=!]{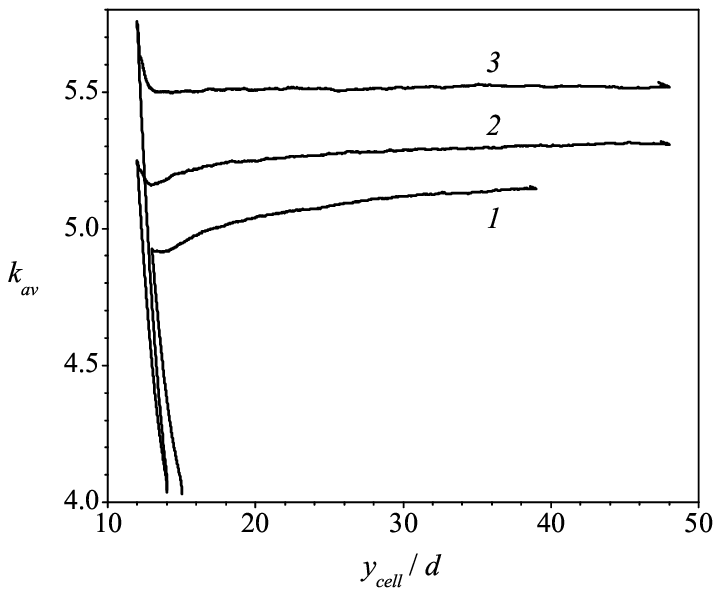}  \hfill
\caption{ \label{f.Gtaux}
  Stress deviator intensity versus model cell size $y_{cell}$ in the simulation
  of shear deformation for the system II with parameter sets (see Table 1)
  No. 7 /line 1/, 3 /2/ and 4 /3/.}
\caption{ \label{f.Gkrox}
  Average coordination number versus model cell size $y_{cell}$ in the simulation
  of shear deformation for the system II.
  Lines are marked by the same symbols as in Fig. \ref{f.Gtaux}.}
\end{figure}

Figures \ref{f.Gtaux} and \ref{f.Gkrox} demonstrate the behavior of selected properties
  of the model cell (the stress deviator intensity and the average coordination number
  for system II) versus the cell size $y_{cell}$.
As can be seen, for the parameter set No. 7 (lines 1) the duration of the simulation process,
  i.e. value $n_z$, is insufficient for the foolproof reaching of the critical (stationary)
  state.
It should be noted that this set corresponds to the density $\rho=39$\%.
Density increase contributes to faster reaching of the stationary state.
For instance, in the case of set No. 4 (lines 3 in the figures) $\rho=51$\% and the state
  of the model cell reaches constant values almost immediately.

\begin{figure}
\centering
\includegraphics*[width=0.49\textwidth, height=!]{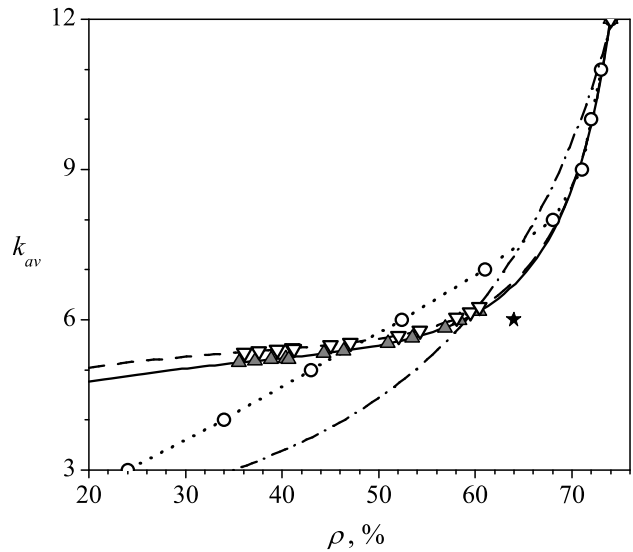}  
\caption{ \label{f.Gkrosdv}
  Average coordination number $k_{av}$ versus density under simple shear deformation.
  Triangles are the simulation data for model systems I (light) and II (dark);
  circles are the data for the regular packing \cite{l.Povar};
  asterisk is the RCP state \cite{l.Zhu08};
  dot-and-dash line is relation (\ref{kroCast}) presented in Ref.~\cite{l.Castel}.}
\end{figure}

To determinate the values characterizing the critical state, the dependences on $y_{cell}$
  of all analyzing parameters at range of $y_{cell}>20d$ were approximated by the expression
  $f(y_{cell}) = f_c + f_1 y_{cell}^{-2} + f_2 y_{cell}^{-3}$.
Figure \ref{f.Gkrosdv} presents the obtained values of the average coordination number
  $k_{av}$ versus powder body density in the critical state.
Calculated points for both model systems is approximated by the dependence
\begin{equation}
k_{av} = k_{av,1} + k_{av,2} \rho + k_{av,3} \rho^2 + \frac{k_{av,4}}{\rho_{*}-\rho}~,
\label{kroapp}
\end{equation}
  with coefficients: $k_{av,1}=3.326$, $k_{av,2}=4.145$ (system I);
  $k_{av,1}=2.943$, $k_{av,2}=4.648$ (system II);
  $k_{av,3}=-8.888$, $k_{av,4}=0.765$, $\rho_{*}=0.813$.
In addition to the calculated points, the condition $k_{av}(0.74)=12$ was assumed for
  the approximations (\ref{kroapp}).
Besides our calculated data, the data of regular 3D-structure density \cite{l.Povar},
  the RCP parameters \cite{l.Zhu08}, and the relation 
\begin{equation}
k_{av} = \frac{\pi/2}{(1-\rho)^{3/2}}
\label{kroCast}
\end{equation}
  proposed in Ref.~\cite{l.Castel} are given in the same figure for comparison.
The figure shows that in examined density range (30-60\%) the calculated values of
  coordination numbers rather weakly depend on density in contrast to
  regular structure coordination numbers or the relation (\ref{kroCast}).
The dependences $k_{av}(\rho)$ of two model systems are close to each other but difference
  between them (about 0.2) exceeds the calculated error by an order.
At that, system I (without hard bonds) is characterized by larger values of the coordination
  numbers.

\begin{figure}[tb]
\centering
\includegraphics*[width=0.99\textwidth, height=!]{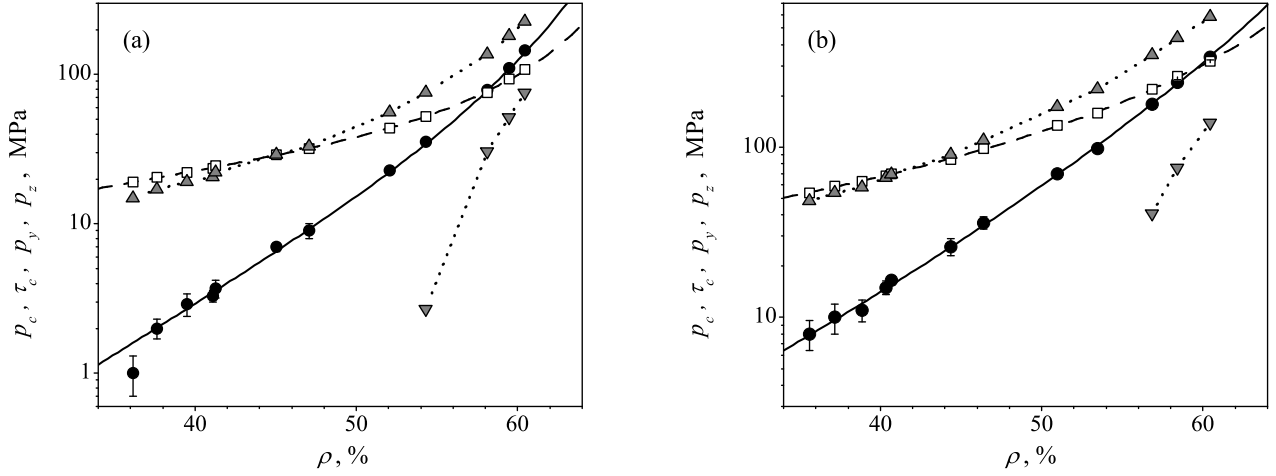}
\caption{
  Stress deviator intensity $\tau_c$ (light squares, dashed lines), hydrostatic
  pressure $p_c$ (dark circles, solid lines), minimum pressure $p_y=-\sigma_{yy}$ (down
  triangles), and maximum pressure $p_z=-\sigma_{zz}$ (up triangles) versus powder body
  density under simple shear deformation for the system I (a) and II (b).
  Points are the simulation data, lines for $\tau_c$ and $p_c$ are the approximations
  by Eq.~(\ref{ptauapp}).}
\label{f.Gtausdv}
\end{figure}

Figure \ref{f.Gtausdv} presents the critical (stationary) values of the stress deviator
  intensity $\tau_c$, the hydrostatic pressure $p_c$, and the minimum and maximum components
  of the stress tensor ($p_y=-\sigma_{yy}$ and $p_z=-\sigma_{zz}$) versus the density of
  the examined systems.
Since logarithmic scale is used along the ordinate axis, the minimum stress components
  ($p_y$) are shown only for large densities where they have positive values.
The same as for the average coordination number, the calculation points in
  Fig. \ref{f.Gtausdv} are fitted satisfactorily by the common curves.
This indicates that the realized algorithm allows us to obtain the critical state reliable
  enough.
The critical state is only determined by the model system density and does not depend on such
  calculation parameters as the particle number $N_p$, the model cell size $L_0$,
  and the duration of the simulation process $n_z$.
The simulation results for the invariants $\tau_c$ and $p_c$ in Fig. \ref{f.Gtausdv} are
  approximated by expressions 

\begin{equation}
\ln\left(\tau_c/p_*\right) = \tau_1 + \tau_2 \rho + \tau_3 \rho^2
                 + \frac{\tau_4}{\rho_{lim}-\rho}~, \ \ \ \ \
\ln\left(p_c/p_*\right) = p_1 + p_2 \rho + p_3 \rho^2 + \frac{p_4}{\rho_{lim}-\rho}~,
\label{ptauapp}
\end{equation}
  where $\tau_1=-0.533$, $\tau_2=7.966$, $\tau_3=-10.951$, $\tau_4=0.916$,
  $p_1=-6.637$, $p_2=17.525$, $p_3=-8.296$, $p_4=0.837$ (for the system I);
  $\tau_1=2.649$, $\tau_2=0.666$, $\tau_3=3.379$, $\tau_4=0.310$,
  $p_1=-2.102$, $p_2=8.963$, $p_3=4.393$, $p_4=0.190$ (for the system II);
  and $p_*=1$ MPa, $\rho_{lim}=0.813$.
Fig. \ref{f.Gtausdv} shows that the examined structures (oxide nanopowders) have
  noticeable positive dilatancy: they tend to increase their volume under simple shear stress.
In our case, it results in positive values of the hydrostatic pressure.
As the density grows the dilatancy increases too.
Whereas the pressure $p_c$ is significantly lower than the tangential stress level $\tau_c$
   at densities $\rho<50$\%, we have already $p_c>\tau_c$ at high densities ($\rho>60$\%). 

Comparison the ratio $p_c/\tau_c$ in systems I and II shows that the creation of hard
  interparticle bonds notably increases dilatancy in the low density region.
For instance, at $\rho\simeq 36$\% this ratio in system I is 0.05 while in system II
  it is almost three times larger (0.14).
However, as the density increases, the values of $p_c/\tau_c$ in systems I and II
  are rapidly move to each other and become equal (about 0.7) at $\rho=55$\%.
At larger densities the system I (without hard bonds) already demonstrates larger values of
  the ratio $p_c/\tau_c$ in comparison to the system II.
While absolute values of the hydrostatic pressure in the system I remains smaller than
  in the system II.

Maximum components of the stress tensor ($p_z$) represented in Fig. \ref{f.Gtausdv} are
  significantly lower than the axial pressures ($p_{out}=p_z$) in the processes A--D
  analyzed in the previous section.
Evidently, in both systems (I and II) the increase of the shear deformation contribution makes
  the powder compaction easier.
For example, at the density $\rho_u=60$\% the simulation results for the examined processes
  are $p_z=210$ /process E/, 480 /D/, 690 /C/, 830 /B/, 1010 /A/ (in MPa, system I);
  $p_z=550$ /E/, 1200 /D/, 1600 /C/, 1750 /B/, 1830 /A/ (in MPa, system II).
The pressures $p_z$, which are necessary for shear deformation initiation in the E-process,
  could be considered as a lower limit of the external forces required for reaching
  the assigned density.

\vspace{5mm}

{\bf V. Loading surface}

The loading surface is a key parameter of the powder body for describing its mechanical
  properties within continuum approach.
It determines the boundary of initiation the plastic-irreversible deformation processes.
In contrast to plastically incompressible materials, in particular, compact metals,
  the powder loading surface is determined not only by the stress deviator intensity ($\tau$)
  but also the first invariant of the stress tensor ($p$).
Moreover, its position in the $p-\tau$ plane depends on the current density (or the porosity
  $\theta=1-\rho$) and tends to the yield condition of solid material at $\theta\to 0$.
To describe the behavior of porous bodies, which reaction to the change
  of the load sign can be neglected, many researchers use the approximation of the loading
  surface in the form of \cite{l.Shtern,l.Max08,l.Olev04}:
\begin{equation}
\frac{p^2}{\Psi(\theta)} + \frac{\tau^2}{\phi(\theta)} = (1-\theta) \tau_0^2(\Gamma_0)~,
\label{loadelli}
\end{equation}
  where $\tau_0$ is the yield stress of solid phase,
  $\Gamma_0$ is the effective shear strain in it,
\begin{equation}
\Gamma_0 = \int \gamma_0 \, dt~, \ \ \ \ \
\gamma_0^2 (1-\theta) = \Psi(\theta) \, e^2 + \phi(\theta) \, \gamma^2~,
\label{Gamma0}
\end{equation}
  $e$ and $\gamma$ are the trace and the deviator intensity of the macroscopic strain rate
  tensor, respectively, and the porous functions $\Psi$ and $\phi$ are defined within the
  hydrodynamic analogy of the elasticity theory \cite{l.Skor,l.ZTF07}.
The hardening rule $\tau_0(\Gamma_0)$ should be determined empirically, for instance, by
  the experimental curves of uniaxial compression.
While stresses reach the surface (\ref{loadelli}), the deformation behavior is defined by
  the associated rule.
It requires the orthogonality of the strain rate vector to the loading surface in the stress
  space \cite{l.Shtern}.
This leads to the requirement of deviator coaxiality for the tensors $e_{ij}$ and $\sigma_{ij}$
  and, with regard to the surface (\ref{loadelli}), to the scalar relation
\begin{equation}
\Psi e \tau = - \phi \gamma p~.
\label{scalar}
\end{equation}
For a given deformation conditions the relations (\ref{loadelli}) and (\ref{scalar}) allow us
  to express any of the stress tensor components as a function of density (porosity).
In the case of analyzed processes for axial pressure $p_{out}=-\sigma_{zz}$ we have
\begin{equation}
\mathbf A: \ \ \ \
p_{out} = \sqrt{\Psi}\ \sqrt{1-\theta} \ \tau_0(\Gamma_0)~, \ \ \ \ \
\Gamma_0 = \int_\theta^{\theta_0} \sqrt{\Psi} \frac{d\theta}{(1-\theta)^{3/2}}~,
\label{AA}
\end{equation}
\begin{equation}
\mathbf B: \ \ \ \
p_{out} = \sqrt{\Psi+\frac16\phi}\ \sqrt{1-\theta} \ \tau_0(\Gamma_0)~, \ \ \ \ \
  \Gamma_0 = \int_\theta^{\theta_0} \sqrt{\Psi+\frac16\phi} \frac{d\theta}{(1-\theta)^{3/2}}~,
\label{BB}
\end{equation}
\begin{equation}
\mathbf C: \ \ \ \
p_{out} = \sqrt{\Psi+\frac23\phi}\ \sqrt{1-\theta} \ \tau_0(\Gamma_0)~, \ \ \ \ \
  \Gamma_0 = \int_\theta^{\theta_0} \sqrt{\Psi+\frac23\phi} \frac{d\theta}{(1-\theta)^{3/2}}~,
\label{CC}
\end{equation}
\begin{equation}
\mathbf D: \ \ \ \
p_{out} = \left(\Psi+\frac53\phi\right) \sqrt{\frac{3(1-\theta)}{3\Psi+14\phi}} \ \tau_0(\Gamma_0)~, \ \ \ \ \
  \Gamma_0 = \int_\theta^{\theta_0} \sqrt{\Psi+\frac{14}{3}\phi} \frac{d\theta}{(1-\theta)^{3/2}}~.
\label{DD}
\end{equation}
This expressions show that the relation $p_{out}(\theta)$ should significantly depend on
  the pressure conditions.
Estimations, represented in Ref.~\cite{l.PRE13}, show that the definition of the hardening
  rule $\tau_0(\Gamma_0)$ on the base of the uniaxial compression curve (process C) gives density
  differences about 10\% according to the equations (\ref{AA}) and (\ref{CC}) for the system I
  at the pressure $p_{out}=100$ MPa.
The calculated curves in Fig. \ref{f.Group} demonstrate at this pressure one order less
  difference on the density.
It is about 1\%.

Inapplicability of the loading surface (\ref{loadelli}) to the systems examined is also
  conformed by the dilatancy effect detected in the previous section.
Symmetry of the surface (\ref{loadelli}) relative to the deviator axis presupposes
  absence of the dilatancy.
In particular, Eq.~(\ref{scalar}) shows that the value $p$ should become zero
  under the condition $e=0$ (shear deformation without compaction, E process).

These evidences identify the difference between powder and porous bodies:
  the powder relatively weakly resists the tensile strains.
As a result, the loading surface locates in the right part of the $p-\tau$ plane.
In contrast to the powder, the sintered porous body is characterized by presence of the
  created interparticle contacts, which leads to its equal resistance to the tensile and
  compression strains.
Thus, identified incompatibility of the mechanical properties of the modeled powder systems
  with the loading surface (\ref{loadelli}) makes necessary to use more complex forms
  of the loading surface.

\begin{figure}  \hfill
\includegraphics*[width=0.49\textwidth, height=!]{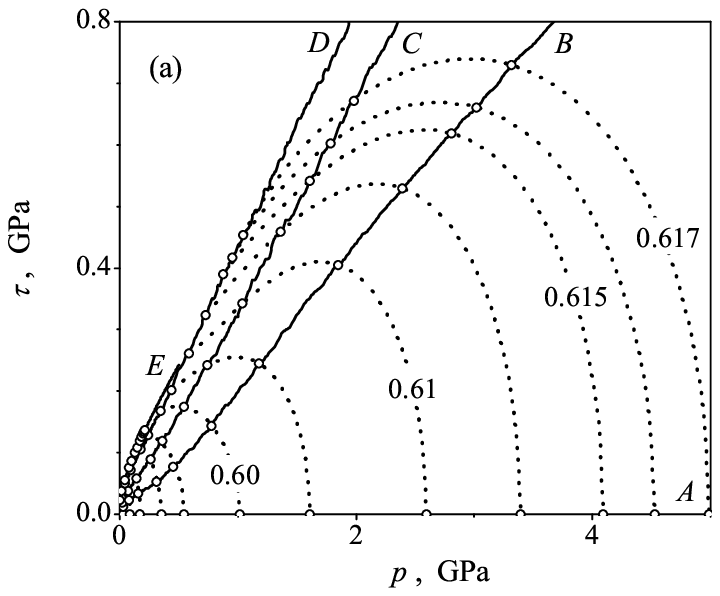} \hfill
\includegraphics*[width=0.49\textwidth, height=!]{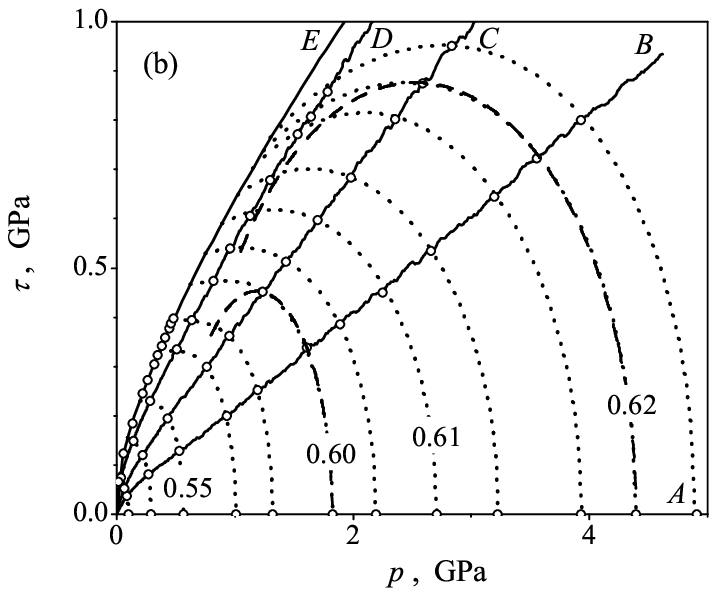} \hfill
\caption{ \label{f.Gtaup}
  Stress deviator intensity versus hydrostatic pressure for the system I
  (a) and II (b).
  Solid lines: monotonic loading curves for the processes A--D (line A coincides with
  the abscissa axis) and line $\tau_c(p_c)$ of the E process.
  Dotted lines: loading surface (\ref{load02}) at constant density values $\rho_u=40$,
  50, 55, 58, 59, 60, 60.5, 61\% (in both systems), and also 61.3, 61.5, 61.6, 61.7\%
  for the system I and 61.4, 61.8, 62, 62.2\% for the system II.
  Dashed lines for the system II are loading surfaces (\ref{load01}) at values
  $\rho_u=60$ and 62\%. }
\end{figure}

Figure \ref{f.Gtaup} demonstrates the monotonic loading curves of systems I and II,
  corresponding to the modeled processes A--D, in the space of the stress tensor
  invariants and the dependences $\tau_c(p_c)$ characterizing the shear deformation
  process (E).
The states corresponding to the constant values of the unloading density $\rho_u$ are
  marked by the points on the curves.
We can see that position of these points requires at least displacement of the loading
  ellipse towards positive values of $p$.
Therefore, we can represent the loading surface equation in the form of
\begin{equation}
(p-p_0)^2 + \tau^2 q_1(\theta) = q_2(\theta,\Gamma_0)~.
\label{load01}
\end{equation}
As a first approximation, we can assume that all three unknown parameters ($p_0$, $q_1$
  and $q_2$) depend only on the porosity.
Then it is not a problem to determine them from any three points corresponding to the
  constant porosity values.
Loading surfaces constructed in this way for system II at density values $\rho_u=0.60$
  and 0.62 are shown in Fig.~\ref{f.Gtaup} to the right (dashed lines).
The points of the processes A, B, and C are used for their construction.
We can see that continuations of the constructed ellipses do not reach the corresponding
  points on the lines of D and E processes.
This can be connected with the neglect of the dependence of the parameter $q_2$ on the
  accumulated effective strain $\Gamma_0$.
The transition from the process A (uniform compression) to the curves $\tau(p)$ of the
  processes B--E along the constant density lines corresponds to the increase of the shear
  deformation contribution.
This should lead to growth of the values $\Gamma_0$ and, therefore, increase the ellipse
  "dimension".
Thus, taking into account the influence of the accumulated strain ($\Gamma_0$), we can
  describe all represented processes by the "shifted ellipse"{} surface (\ref{load01}).

With the aim of indirect taking into account the dependence of the parameter $q_2$ on the
  compaction conditions, the following approximation of loading surface was applied
\begin{equation}
(p-p_0)^2 + \tau^2 q_1(\theta) = \left(p_A-p_0\right)^2 \left[ 1 + \alpha \left(p_A - p\right)^m \right]~.
\label{load02}
\end{equation}
Here $p_A$ is the value of $p$ on the A-line, that is the maximum value $p$ for a given density;
  $p_0$, $q_1$, $\alpha$, and $m$ are the parameters depending on the density (porosity).
This approximation due to the second term in the square brackets takes into account
  increasing of the ellipse "dimensions"{} during transition from the A-process to B, C
  and so on.
The results of applying the formula (\ref{load02}) are represented in Fig. \ref{f.Gtaup}
  by dotted lines.
It was accepted in this case for all density values that $p_0=0.7p_A$ and remaining parameters
  were determined from the condition of the best describing the points B, C, D and E.
Figure \ref{f.Gtaup} shows that the approximation (\ref{load02}) allows us to reproduce
  accurately enough all calculation points besides the line E (simple shear).
Before passing through the corresponding point on the E-process curve, the constant density
  line crosses the curve E in the region of larger density values.
The latter, in our view, indicates the end of the loading surface in the E-line neighborhood.
If we neglect the strain hardening contribution to the strain resistance of the examined
  powder systems, the E-line can be associated with the fracture surface of the model systems.
This qualitatively corresponds with the data \cite{l.Piz} on this boundary location.
It should be noted that the E-line in the system I closely coincides with the D-process curve.
Therefore constructed loading surfaces end at the D-line.

\vspace{5mm}

{\bf VI. Associated flow rule}

The loading surface, constructed in the previous section, allows us to test the
  applicability of the associated flow rule to description of the nanopowder behavior.
One implication of the associated flow rule is the deviator coaxiality for the stress
  and strain rate tensors, that is $\tau_{ij}\propto\gamma_{ij}$.
It is not difficult to determine that in the cases of A, B, and C processes the feasibility
  of the coaxiality is ensured by the symmetry condition, i.e., the equality of stresses
  along the directions with equal strain rates.
This fact is not difficult to prove in common case.
Let principal axes of the both tensors coincide (that is, in our case, the material
  is isotropic) and two diagonal elements are equal.
Let’s take for definiteness that $\sigma_1=\sigma_2\not=\sigma_3$ and $e_1=e_2\not=e_3$.
Then the deviator components of these tensors can be found as $\tau_1=\tau_2=-\tau_3/2$
  and $\gamma_1=\gamma_2=-\gamma_3/2$.
That means that the deviators are coaxial.

\begin{figure}  \hfill
\includegraphics*[width=0.49\textwidth, height=!]{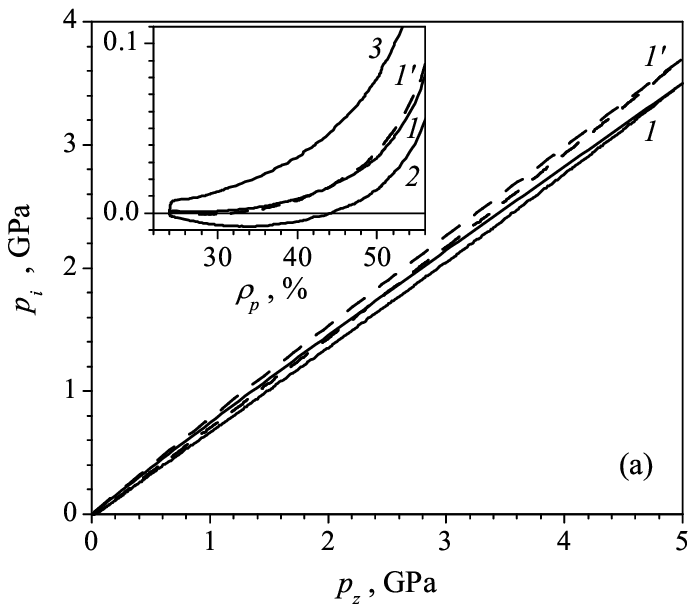} \hfill
\includegraphics*[width=0.49\textwidth, height=!]{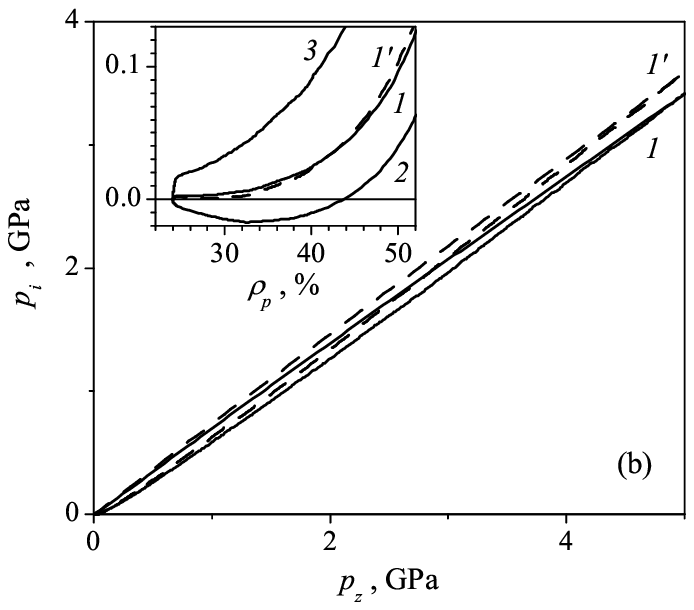} \hfill
\caption{ \label{f.GassD}
  Calculated values of the stresses along $Ox$ axis ($p_x=-\sigma_{xx}$, solid lines
  {\textit 1}) and values of $p_x^{(as)}$ corresponding to the associated flow rule
  (Eq.~(\ref{passo}), dashed lines {\textit 1}$\,'$) for the process D in the systems I
  (a) and II (b).
  The lines of monotonic loading up to $p_z=5$ GPa and elastic unloading are shown.
  The insets show the values $p_x$ /1/, $p_x^{(as)}$ /{\textit 1}$\,'$/, $p_y$ /{\textit 2}/,
  and $p_z$ /{\textit 3}/ versus density $\rho_p$ in the low pressure zone.}
\end{figure}

In the cases of D (compression with shear deformation) and E (simple shear) processes
  the coaxiality of the stress deviator to the strain rate deviator leads to the need
  of following relations between stress tensor components:
\begin{equation}
\mathbf D: \  p_x^{(as)} = (2p_y+p_z)/3~; \ \
\mathbf E: \  p_x^{(as)} = (p_y+p_z)/2~.
\label{passo}
\end{equation}
Figure \ref{f.GassD} demonstrates rather good coincidence of the calculated values $p_x$
  with the value $p_x^{(as)}$ in the D-process case.
This can be considered as the confirmation of the associated flow rule.
For instance, at $p_z=5$ GPa the difference between $p_x$ and $p_x^{(as)}$ is about 6\%
  in both systems.
This difference in the whole pressure range is significantly smaller than differences between
  separate components of the stress tensor (see insets in Fig. \ref{f.GassD}).
In the case of E-process the difference between $p_x$ and $p_x^{(as)}$ is rather larger:
  it reaches 14\% from the difference $p_z-p_y$.
This fact can be related to the particular status of the E-process that was discussed
  in the previous section, namely, closeness or even identity of the E-process with
  the fracture surface of powder body. 

\begin{figure}[tb]
\centering
\includegraphics*[width=0.98\textwidth, height=!]{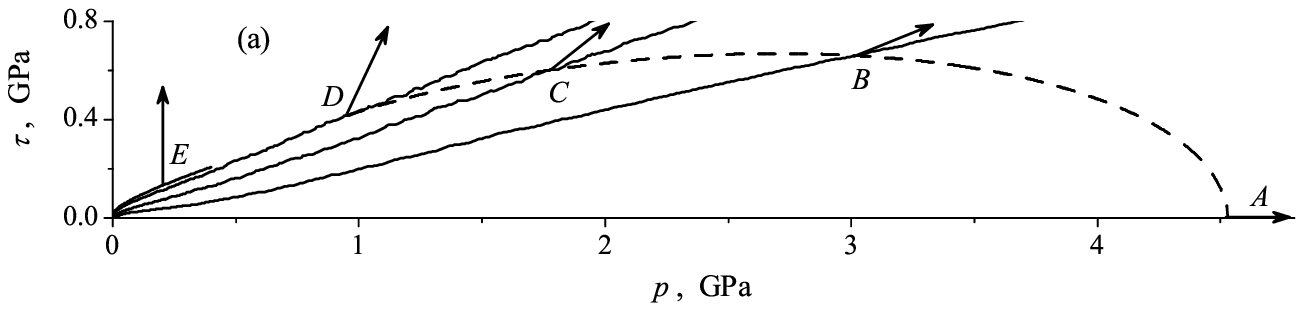}
\includegraphics*[width=0.98\textwidth, height=!]{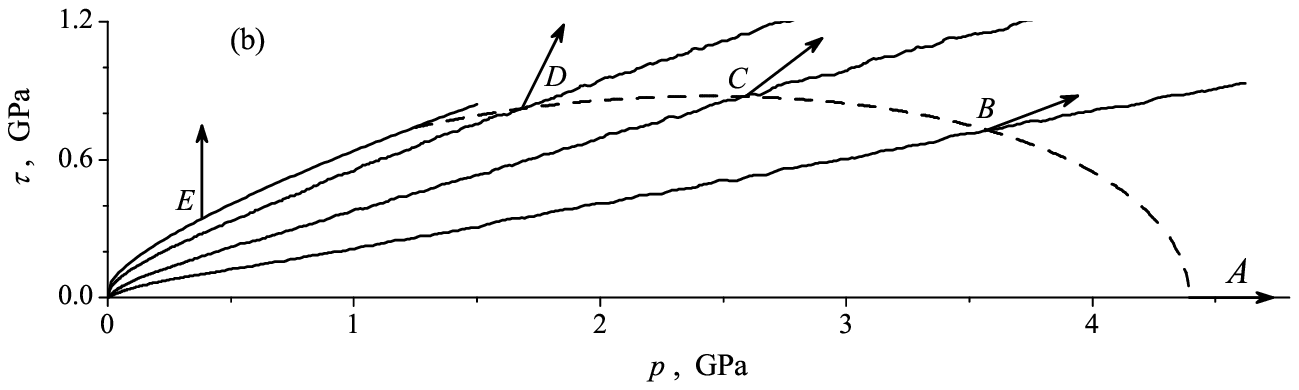}  
\caption{Stress deviator intensity versus hydrostatic pressure for systems I (above)
  and II (below). Solid lines correspond to the processes A--E (line A coincides with
  the abscissa axis), dashed line is the loading surface (\ref{load02}) at value
  $\rho_u=61.6$\% in the system I and 62.0\% in the system II.
  Arrows represent the vectors ($e$, $\gamma$) defining strain "direction"{} in the
  processes A--E (here the positive value $e$ corresponds to the compression).}
\label{f.Gorto}
\end{figure}

Another implication of the associated flow rule, in addition to the coaxiality of the tensors
  $\tau_{ij}$ and $\gamma_{ij}$, is the orthogonality of the vector ($e$,$\gamma$),
  which specifies the "direction"{} of the deformation process, to the loading surface
  in the $p-\tau$ plane.
In particular, it leads to the scalar relation (\ref{scalar}) for the elliptic-type
  surface (\ref{loadelli}).
Figure \ref{f.Gorto} presents the loading surfaces corresponding to the densities
  $\rho=61.6$\% and 62.0\% for the systems I and II, respectively, and also
  the vectors ($e$,$\gamma$) characterizing the A--E processes.
We can see that the orthogonality to the loading surface is observed only in the trivial case
  of the uniform compression (the line of A-process, which coincides with the abscissa axis).
All other vectors noticeably deviate from the normal to larger values $e$, that is,
  the material demonstrates too high densification rate in comparison with the associated
  flow rule for given stresses.
For example, the positions of the points C and D on the loading surface of I system and
  the points D on the loading surface of II system indicate a need for density decrease
  in this processes.
This fundamentally contradicts to the conducted numerical experiments where the processes
  C and D have corresponded to the density increase.

Thus, we can establish the nonapplicability of the associated flow rule to the mechanical
  properties description of the oxide nanopowders due to the failure of one of its
  implications (the orthogonality of the vector ($e$,$\gamma$) to the loading surface).
The non-orthogonality of the vectors ($e$,$\gamma$) to the loading surface is
  observed more appreciable for the system I which has no hard interparticle bonds.
Appearance of such bonds in the system II makes it more similar to the behavior of a
  plastic porous body.
This deviation from the orthogonality principle occurs rather often in the mechanics of
  granulated solids and soils.
Therefore, a number of authors prefer to introduce the loading surface and the plastic
  potential independently of each other \cite{l.Rudni,l.Gara,l.Holc}.
The need for the plastic potential introduction is explained by the fact that it 
  determines the relations between stresses and strain rates in the plastic material.
This provides the possibility to formulate and solve initial boundary value problems.
Furthermore, the presence of potential ensures the variational principle implementation
  that simplifies significantly numerical solution procedures.

Factors contributing to the non-orthogonality of the vectors ($e$,$\gamma$) to the loading
  surface include, first of all, lack of comprehensive information on tolerance for
  the plastic strain.
Here it should be taking into account that the energy dissipation in the examined systems
  has a very low threshold.
Therefore, the insignificant residual strains in powder systems as well as in granulated
  solids and soils are already observed on the prime deformation stage.
Moreover, the relative irreversibility threshold can be sensitive to the loading path,
  as experimental data show.

\vspace{5mm}

{\bf VII. Conclusion}

The main results of the study are as follows.

1. The two granular systems corresponding oxide nanopowders have been examined by the
     granular dynamics method.
   The I system, which has no hard interparticle bonds of chemical nature, corresponds
     to the non-agglomerating nanopowders because of layers of adsorbed gases on the particles.
   The II system, which includes creation and destruction of hard bonds, corresponds to the
     purified nanopowders strongly inclined to agglomeration.
   For these systems the quasistatic processes of the uniaxial, biaxial, uniform compression
     and the compression with shear deformation of the model cell have been simulated.
   The elasto-reversible contribution mainly connecting with the elastic strain of individual
     particles and the plastic irreversible contribution connecting with mutual displacements of
     particles have been extracted from the total deformation.
   The plastic part of the deformation is characterized by the unloading density $\rho_u$.
   It has been determined that in the coordinates "$\ln(p_{out})-\rho_u$", where $p_{out}$
     is the maximal external pressure, the curves of all examined processes are close to
     each other for densification from the equal initial density ($\rho_0=24$\%).
   The density differences are less than 2\%.
   The final powder state in the high pressure range is close to the parameters of
     random close packing, i.e., the density $\rho\simeq 64$\%, the coordination number
     $k\simeq 6$.
   On the compaction curves $p_{out}(\rho_u)$ we can identify three stages:
   1) the stage of insignificant density change with pressure increasing,
      when the external loading is insufficient to overcome the initial particle bonds;
   2) the stage of "logarithmic"{} densification ($\Delta\rho\propto \ln(p_{out})$),
      when the intensive particle rearrangement occurs;
   3) the stage when the density tends to some maximum value $\rho_{\max}$.
   For third stage the dependence $p_{out}\propto 1/(\rho_{\max}-\rho)^{D-1}$
     has been established, where $D$ is the space dimension.

2. The oxide nanopowder behavior under shear deformation has been simulated.
   Here the effect of the positive dilatancy has been found out: the examined systems
     tend to increase their volume under the simple shear stress.
   In simulation results (at constant volume) this fact appears as positive values of the
     hydrostatic pressure $p_c$.
   As the density grows the dilatancy increases too.
   Whereas the pressure $p_c$ is significantly lower than the tangential stress level $\tau_c$
     at densities $\rho<50$\%, we have already $p_c>\tau_c$ at high densities ($\rho>60$\%). 
   Comparison of examined systems I and II shows that the appearance of hard bonds
     significantly increases the dilatancy in the low density range ($\rho<50$\%).
   But qualitatively the properties of both systems are similar. 
   This fact allows us to conclude that the main reason of the positive dilatancy
     in whole investigated density range is the high dispersive attraction forces
     inherent to nanopowders.

3. The approximating formula for the loading surfaces of the examined systems
     has been offered.
   It has been established that these surfaces are close to the elliptic-type surface.
   However, the loading ellipse in the $p-\tau$ plane is strongly shifted towards
     positive values of $p$ and its shape is determined by not only the powder porosity
     but also the specific features of loading.
   The suggested approximation of the loading surface allows us to reproduce satisfactorily
     all examined processes except for the shear deformation of the model cell (E-process).
   It has been identified that before passing through the corresponding point on the E-process
     curve (in the $p-\tau$ plane) the constant density line crosses the curve E in the region
     of larger density values.
   The latter can indicate the end of the loading surface in the E-line neighborhood.
   Based on this fact we can conclude that the E-process curve can be associated with
     the fracture surface of the systems examined.

4. The nonapplicability of the associated flow rule to the mechanical properties description
     of the oxide nanopowders has been found out.
   At the same time, one implication of the associated flow rule, that is the deviator
     coaxiality of the stress and strain rate tensors, remains valid.
   But another implication, that is the orthogonality of the vectors ($e$,$\gamma$) to
     the loading surface in the $p-\tau$ plane, is broken.
   This fact indicates that the loading surface for the oxide nanopowders does not coincide
     with the plastic potential isolevels and, therefore, cannot be used for the estimation
     of the system dissipative behavior. 

The work is supported by the RFBR, projects 12-08-00298 and 14-08-90404.

%\clearpage

\end{document}